# Extending Message Passing Interface Windows to Storage

Sergio Rivas-Gomez, Stefano Markidis, Ivy Bo Peng, Erwin Laure
Department of Computational Science and Technology
KTH Royal Institute of Technology
Stockholm, Sweden
{sergiorg, markidis, bopeng, erwinl}@kth.se

Gokcen Kestor, Roberto Gioiosa
Computational Science and Mathematics Division
Pacific Northwest National Laboratory
Washington, USA
{gokcen.kestor, roberto.gioiosa}@pnnl.gov

*Abstract:* **This work presents an extension to MPI supporting the one-sided communication model and window allocations in storage. Our design transparently integrates with the current MPI implementations, enabling applications to target MPI windows in storage, memory or both simultaneously, without major modifications. Initial performance results demonstrate that the presented MPI window extension could potentially be helpful for a wide-range of use-cases and with low-overhead.**

## I. Introduction

As high-performance computing systems grow in size and complexity, the amount of data consumed and generated dramatically increases as well. With the first wave of systems breaking the ExaFLOP barrier expected to appear during the 2020-decade timeframe, it is likely that the next Exascale milestone will be Exabyte (i.e., billions of Gigabytes per second produced by a large magnitude of calculations). Even assuming a constant amount of data produced per thread, the number of concurrent threads conducting I/O operations at Exascale will be much larger than today.

In this regard, the trend for large-scale computer design is diverging from the traditional compute-only node approach, that uses a separate parallel storage system, to hybrid solutions where data is moved next to the computation. For instance, Summit, the next Supercomputer from Oak Ridge National Laboratory, will combine DRAM with 800GB of non-volatile RAM (NVRAM) per node [1].

Despite the opportunity that these changes may bring, handling data efficiently and transparently while maintaining the current programming models is difficult to accomplish. Nonetheless, these programming models need to evolve to incorporate the upcoming changes in the memory / storage hierarchy [2]. While traditional I/O methods like MPI IO can be used to address storage, we believe that a tighter integration in the memory management of the application will likely provide more flexibility and performance advantages.

In this work, we extend the MPI one-sided communication model to support window allocations in storage and together with traditional memory-based allocations. Our objective is to define a seamless extension to MPI that could provide benefits for current and future storage technologies without altering the MPI standard, allowing to target either files (i.e., for local and remote storage through a parallel file system) or alternatively address block devices directly (i.e., as in DRAM).

From a programmer standpoint, applications benefit from this feature transparently without major modifications. Hence, our approach can be applied to novel use-cases, for instance:

- *Transparent persistence* - Exposing part of the storage for one-sided communications imply that changes are transparently synchronized in storage, allowing for tight coupling scenarios.
- *Fault tolerance* - Having the exposed windows synchronized to storage enables implicit resilience. If a process / node fails, it is theoretically as simple as mapping once again the part of the storage missing.
- *Big Data applications* - Data analytics require large input data sets that produce large amounts of output data. Exposing part of the storage in a one-sided model can be beneficial and simplify cooperation between processes.

Initial performance experiments reveal that the potential of having storage exposed as part of the global address space is beneficial. With optimizations, only a 10% performance degradation is ideally observed when running the STREAM benchmark using 1000 million elements in local storage mode compared to the traditional memory-based approach.

This paper is organized as follows. Section II describes the one-sided communication model of MPI, together with how the MPI window extension has been designed and integrated following features of the MPI standard. Section III presents performance results mostly based on a modified version of the STREAM library. Section IV provides information on related work. Lastly, Section V summarizes our conclusions.

## II. Methodology

MPI is the de-facto standard for programming large-scale distributed-memory systems. Since its inception, MPI has provided high performance, efficiency and portability on massively parallel systems. Together with the traditional message passing model of MPI, MPI-2 additionally included remote memory operations [3]. These operations are meant to provide access to the local memory of other processes, a feature known as the one-sided communication model and that differs from the traditional two-sided counterpart (i.e., of send plus receive operations) by not requiring cooperation between each side of the communication.

Following the one-sided communication model, each process exposes part of its local memory through the concept of an MPI "window". The size of each window can differ considerably among processes, and, as a matter of fact, some may not even share memory if desired. Traditionally, the purpose of MPI windows is to allocate or share part of the processes memory space. However, as high-performance computing systems grow in size and scientific problems become larger, the amount of data handled per application might constrain the possibilities of this model. Furthermore, some of this data may be necessary to be later saved to storage for different purposes, requiring an additional step.

In this section, we present an extension to storage for the MPI one-sided communication model. Processes can, through their window, expose part of their memory, local / remote storage, or both simultaneously. The extension does not require any alteration to the MPI standard and transparently integrates with the current interface specification, enabling existing applications to take advantage of this feature without major modifications.

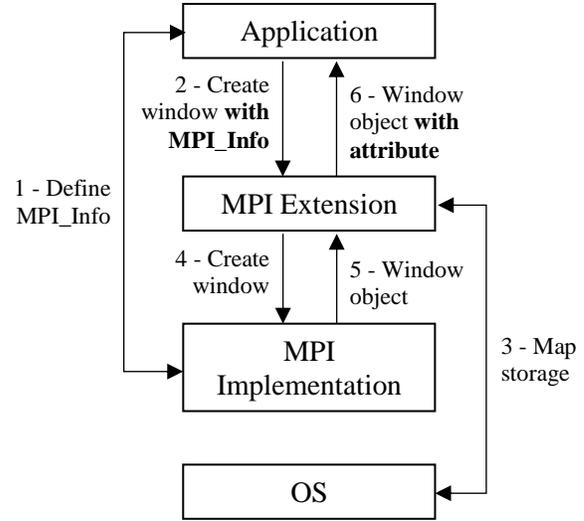

Figure 1. Flow diagram of the allocation of an MPI window in storage.

Compared to POSIX or efficient parallel I/O alternatives such as MPI IO, the approach represents a tighter integration in the memory management of the application and seamlessly supports addressing files and block devices indistinctively (e.g., NVRAM). Moreover, its versatility provides automatic caching advantages and allows for implicit synchronous and asynchronous storage, instead of the traditional explicit I/O model through read / write operations.

### A. Extending the MPI window concept to storage

Our design integrates with MPI without requiring an additional set of operations to support it. This is possible thanks to the versatility of the interface, taking into special consideration the following three aspects:

- *Window allocation* - Allocating a window in MPI can be manually or through the MPI implementation. The former has the benefit of exposing an already existing memory allocation to other processes; while the latter creates a new allocation that is handled by the MPI implementation. We concentrate on this alternative.
- *Performance hints* - This concept of MPI represents a key-value container that allows to alter an MPI implementation (e.g., file stripping size, in MPI IO). The hints are ignored if not supported.

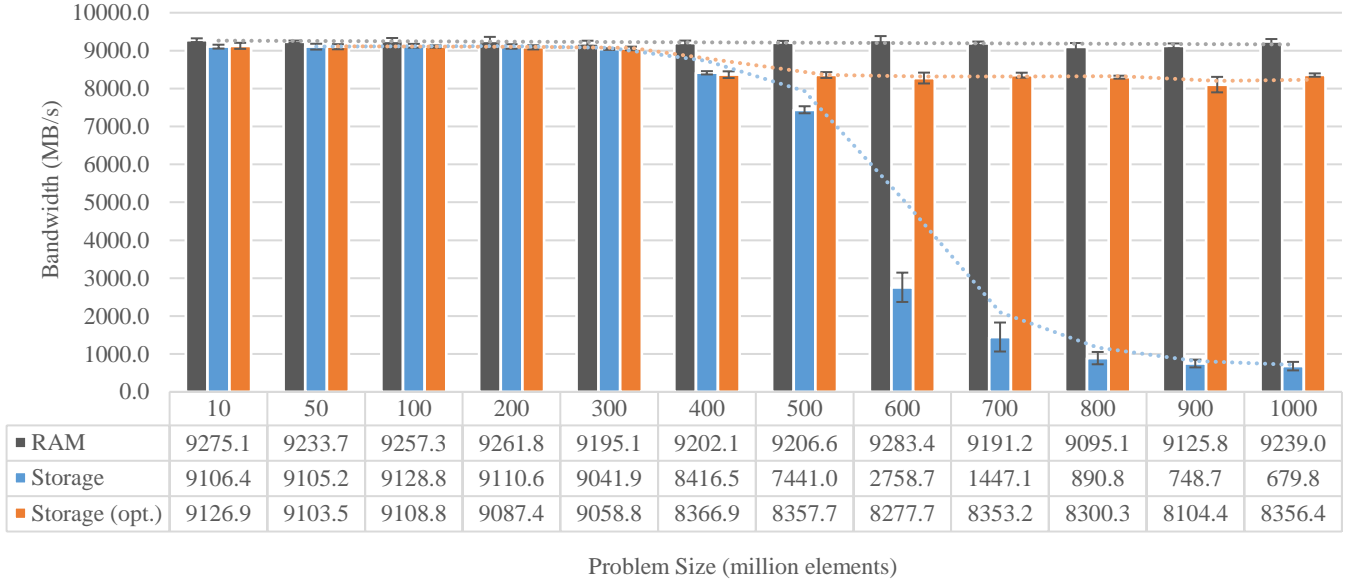

Figure 2. Bandwidth for the MPI one-sided allocations in RAM (dark gray bars), Storage (ligth blue bars) and Storage optimized (orange bars), against a modified version of the STREAM benchmark running on "Blackdog". The x-axis represents the problem size in million of elements per array, while the y-axis represents the bandwidth.

| | 10 | 50 | 100 | 200 | 300 | 400 | 500 | 600 | 700 | 800 | 900 | 1000 |
|---|---|---|---|---|---|---|---|---|---|---|---|---|
| RAM | 9275.1 | 9233.7 | 9257.3 | 9261.8 | 9195.1 | 9202.1 | 9206.6 | 9283.4 | 9191.2 | 9095.1 | 9125.8 | 9239.0 |
| Storage | 9106.4 | 9105.2 | 9128.8 | 9110.6 | 9041.9 | 8416.5 | 7441.0 | 2758.7 | 1447.1 | 890.8 | 748.7 | 679.8 |
| Storage (opt.) | 9126.9 | 9103.5 | 9108.8 | 9087.4 | 9058.8 | 8366.9 | 8357.7 | 8277.7 | 8353.2 | 8300.3 | 8104.4 | 8356.4 |

- *Attribute caching* - MPI provides this feature for library writers. Its purpose is to allow caching of special key-value attributes within certain objects, such as communicators or, in our case, windows.

Combining these three concepts, the proposed MPI window extension has enough flexibility to transparently integrate with any MPI implementation. Our MPI window extension resides between the application and the underlying MPI implementation, intercepting the window allocation calls for further consideration while still allowing us to maintain the original application logic unchanged.

The hints provided during the allocation determine the type of allocation request. From a programmer standpoint, the application still receives a pointer to a recently allocated memory, regardless the type. However, in the storage-based, the difference is that now the memory space is mapped to storage through the page cache of the operating system and further changes will be transparently synchronized.

Figure 1 depicts an example flow diagram representing a window allocation in storage by a certain application process. The call from this process is intercepted and analyzed by our MPI extension to understand if the requested allocation is based on memory or storage. In the former case, the extension will directly rely on the underlying MPI implementation. In the latter case, as shown in Figure 1, a mapping in storage is created through the OS and attached to the window. Regardless of the allocation type, an extra attribute is always cached inside the window object. This is an important aspect of window allocation: the cached attribute is necessary to understand how to deallocate the space conveniently.

The described approach only requires minor changes in the application source code. All the MPI one-sided operations work as expected with either memory, storage or hybrid allocations, and the modified applications will remain correct against MPI implementations that do not support the feature (i.e., the allocation is performed in memory).

### III. EXPERIMENTAL RESULTS

In order to understand the potential performance constraints that the storage allocations introduce into the MPI one-sided communication model, we have conducted our experiments using a modified version of the STREAM benchmark [4]. This benchmark is well- known for measuring memory bandwidth (in MB/s) through the execution of four simple vector kernels: *copy*, *scale*, *add* and *triad*.

In our case, our plan is to analyze the bandwidth produced through a modified MPI version of the STREAM benchmark that uses one-sided allocations, both in memory and in local storage (with files). Our aim is to evaluate the performance in terms of local storage bandwidth against traditional DRAM allocations. Additionally, we compare execution times with explicit I/O through POSIX-like functionality of MPI IO.

All the performance tests have been conducted with the *swap* partition disabled in a local machine available at the Department of Computational Science and Technology at KTH Royal Institute of Technology (Table I). The window mapping in storage has also been defined making sure that none of the pages are copied to / from the swap partition, if any. Finally, the STREAM benchmark has been executed multiple times during different days and timeframes, and on each time several runs have been performed.

TABLE I.  SPECIFICATIONS OF LOCAL MACHINE "BLACKDOG"

| Component | Description |
| --- | --- |
| Processor | 2 × 4 core Xeon E5-2609v2 @ 2.5GHz (10MB L3) |
| Memory | 8 × 8GB DDR3 + 4 × 4GB DDR3 (Total: 72GB) |
| Storage | 2 × WDC WD4000F9YZ-09N20L0 4TB (non-RAID) |
| Software | Ubuntu Server 16.04 / gcc v5.4.0 / MPICH v3.2 |

### A. *Evaluating the performance of local storage*

The first series of performance tests have the purpose of comparing the modified STREAM benchmark using local storage with the bandwidth of the original source code using memory allocations. All the operations are performed through local files and do not involve network, which will give us an idea of how storage allocations could affect a given node.

Figure 2 depicts the average of the performance results obtained, including the standard deviation. As expected, the bandwidth of the modified STREAM benchmark using one-sided allocation in DRAM (label "RAM") is equivalent to the original STREAM implementation. The version based on storage (label "Storage") is also promising, but the results show that it begins to perform worse after around 500 million elements per array. We noted that part of the reason for this effect is the flushing policy of the OS, which blocks further IO requests when the amount of data that must be written to the storage layer is large. Thereafter, when a certain threshold is met, the application is allowed to proceed.

In the Linux Kernel, this behaviour can be configured through the virtual memory *vm* settings. By default, our installation allowed up to a 20% of the memory with dirty pages, and a 10% of memory threshold to begin flushing to disk (i.e., by the *pdflush* process, originally set to wake up every 15 seconds). If at some point the 20% limit is reached, the application is blocked and no IO operation is further allowed, decreasing the performance. Therefore and as can be observed in Figure 2 with label "Storage (opt.)", fine-tuning these settings dramatically increases the performance and provides bandwidth values close to DRAM-based allocations.

Nevertheless, it is important to note that the application is still bounded by the storage bandwidth at some point. Finding a good balance of the flushing policy settings and overlapping as much computations as possible will help the OS to continue flushing the modified pages asynchronously. However, if we often require synchronization points with the storage layer, the performance will considerably decrease.

### B. *Evaluating the performance of explicit IO operations*

The second series of performance tests have the purpose of comparing execution times with explicit I/O through POSIX-like operations of MPI IO. The STREAM benchmark has been adapted to compute each kernel in blocks of 1 million of elements instead, meaning that the code will read a block from storage, operate with it and store back the result, in steps of 1 million elements per kernel. Non-blocking operations have been used in some of the read requests of the MPI IO version, while in the MPI extension-based, a synchronization point has been defined after each block. This will inform the OS to asynchronously flush the changes to storage (as in MPI IO, for fair comparison). The tests have been conducted with the optimal *vm* settings of the previous subsection, and the execution time measured also includes the time to flush the last dirty pages to storage, useful to determine how caching affects the obtained results. This is relevant to consider, as the performance will always be limited at some point by the bandwidth of the storage layer, for instance during window deallocation, when exceeding the amount of DRAM available or when no more computations can be overlapped.

Figure 3 reflects the speedup of the average execution time obtained with the block-based STREAM benchmark. Despite some of the write operations being also cached by the OS in the MPI IO version, the results show that our solution provides performance benefits compared to using explicit I/O.

The execution time is also more reasonable and stable with respect to MPI IO, hence the disruptive speedup variations observed.

The figure additionally depicts the speedup forcing the data to be synchronously flushed to storage after each block, with the purpose of understanding the cache effects to / from storage on both versions. Even though the performance in this case is mainly bounded by the bandwidth of the underlying storage layer, the proposed MPI extension shows performance benefits when compared with the MPI IO version, thanks to avoiding I/O requests (e.g., the output of each kernel is transparently hold within the memory mapping, even if the data is forced to be synchronized with storage).

### C. Additional performance considerations

Despite the bandwidth and access latency differences of the I/O subsystem, we have demonstrated that overlapping computations with storage operations substantially hides the existing constraints of this subsystem. This fact is mainly possible thanks to the page cache of the OS, which allows to temporary hold dirty pages mapped to storage. Hence, the memory acts as another level in the cache hierarchy: in the same way that programmers do not handle data movement between the CPU cache and memory, our approach releases handling data movement to disk.

The importance of fine-tuning the virtual memory settings of the OS is consequently critical in order to achieve high-performance, as already discussed. These settings determine the interval and retention period of the dirty pages stored within the page cache of the OS, which means that increasing the amount of active dirty pages in memory allows to compensate for differences in performance.

We also note that going beyond the RAM limit introduces performance constraints, depending on the application. Targeting more memory than physically available might bound the application performance to the bandwidth of the I/O subsystem. This means that, while smaller tests fit into memory and the performance remains high, increasing the problem size to limits close to the dirty ratio (i.e., the limit to flush dirty pages to storage) begins to produce performance degradations. This is something we already observed on STREAM due to the inherently sequential writes of the benchmark, that do not allow to reuse previously stored data inside the page cache of the OS and continuously accumulates write requests.

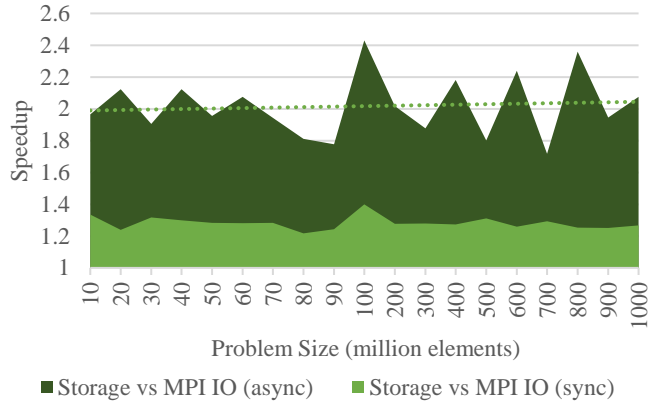

Figure 3. Speedup of the MPI one-sided allocation in Storage versus the equivalent MPI IO implementation of the STREAM benchmark running on "Blackdog". The x-axis shows problem size in million of elements per array, while the y-axis represents the speedup.

## IV. RELATED WORK

Despite being a relatively simple approach, to the best of our knowledge there has not been any published work that directly aims to target transparent one-sided communications over storage without involving changes in the MPI standard.

Bo Peng et al. [5], for instance, defined an effective way to tackle data-intensive applications through a streaming model that uses MPI IO collective operations on dedicated IO nodes. We expect our MPI window extension to provide benefits for similar work, like asynchronous remote storage operations (instead of explicit IO operations).

Other authors have also considered storage as a solution for fault tolerance. The work by Abeyratne et al. [6] is a good example, where a Checkpoint Location Controller (CLC) is defined to offer reliable checkpointing by deciding whether to target memory or per-node storage. Our approach could extend this work by allowing implicit neighbor checkpointing.

Lastly, Dorożyński et al. [7] share some of the concepts presented in this paper. While their aim was mainly to provide checkpointing of parallel applications using non-volatile RAM (NVRAM), our purpose is to demonstrate the possibilities of providing storage support within one-sided communications in MPI, regardless of the use-case.

## V. CONCLUSION

In this paper, we have presented a seamless extension to the MPI one-sided communication model that enables the allocation of MPI windows on storage. The described design provides considerable benefits to applications and without requiring major modifications to the existing logic.

Performance results conducted in a local machine through a modified version of the STREAM benchmark has proved the MPI window extension to be clearly beneficial, even when compared with MPI IO. A fine-tune of the flushing policy settings of the operating system allows low-overhead while still committing changes to storage transparently. From these results, we concluded necessary a comprehensive overlap of the computations in order to achieve high performance.

Lastly, our plan is to analyze the consequences of using remote parallel file systems and direct device mapping in window allocations, as well as to understand the implications of the proposed MPI window extension on real-world HPC applications. The outcome of this work is expected to provide further insights about the overall benefits of the approach.

## Acknowledgment

This work was funded by the European Commission through the SAGE project (Grant agreement no. 671500 / More information at http://www.sagestorage.eu).